\documentclass[fp,twocolumn]{jpsj3}
\usepackage{txfonts}
\usepackage{amsmath,amssymb}
\usepackage{bm}
\usepackage{mathrsfs}
\usepackage{color}
\usepackage{graphicx}

\newcommand{\bra}[1]{\langle #1 |}
\newcommand{\ket}[1]{| #1 \rangle}
\newcommand{\braket}[2]{\langle #1 | #2 \rangle}
\newcommand{\ketbra}[2]{| #1 \rangle\langle #2 |}
\newcommand{\fbra}[1]{\langle\!( #1 |}
\newcommand{\fket}[1]{| #1 )\!\rangle}
\newcommand{\fbraket}[2]{\langle\!( #1 | #2 )\!\rangle}
\newcommand{\fketbra}[2]{| #1 )\!\rangle\langle\!( #2 |}

\title{Selective Remote Dissipation of an Off-resonant State via Indirect Driving}

\author{Hidemasa Yamane$^1$\thanks{yamane.hidemasa@orist.jp}}
\inst{$^1$Osaka Research Institute of Industrial Science and Technology, 2-7-1, Ayumino, Izumi-city, Osaka 594-1157, Japan} 

\abst{We show how local periodic driving can be used to control dissipation in a structured environment in a highly selective manner.
As a minimal setting, we consider two discrete levels coupled to a one-dimensional tight-binding continuum with a finite bandwidth, where only one level is driven while the other remains undriven.
Without driving, both bare energies are placed outside the static continuum band so that neither level decays.
We demonstrate that the drive can nevertheless activate a selective remote dissipation channel: the undriven level acquires a finite decay rate, whereas the driven level can remain long-lived.
The mechanism is clarified within Floquet theory.
Periodic driving generates photon-assisted channels shifted by integer multiples of the drive frequency, effectively creating a ladder of drive-shifted continuum sidebands (Floquet channels).
A decay channel for the undriven level opens once its bare energy overlaps an open sideband accessed via drive-enabled pathways; in the tight-binding example, the decay is strongly enhanced near the sideband edge due to the increased density of states.
The dominant remote pathway is controlled by Bessel-weighted couplings and can be switched and strongly suppressed by tuning the drive amplitude.
We verify these predictions by direct numerical integration of the time-dependent Schr\"odinger equation.
We also formulate a complex-eigenvalue problem for the Floquet Hamiltonian by eliminating the continuum via a Brillouin--Wigner--Feshbach projection, and show that the pole-implied decay rate quantitatively reproduces the time-domain decay envelope.}

\begin{document}
\maketitle

\section{Introduction}

Quantum systems unavoidably couple to their environments and are therefore often described as open systems subject to dissipation and fluctuations. 
In quantum information processing, such unwanted dissipation and decoherence limit gate fidelities and the attainable circuit depth \cite{breuer2002theory, siddiqi2021engineering, blais2021circuit}.
At the same time, controlled system--environment couplings have enabled reservoir-engineering protocols where dissipation becomes a resource for stabilization, initialization, and fast reset \cite{poyatos1996quantum, verstraete2009quantum, shankar2013autonomously, zhou2021rapid}.

A particularly relevant platform for tailoring dissipation is a discrete quantum emitter coupled to a structured one-dimensional continuum, as in waveguide quantum electrodynamics (QED) \cite{sheremet2023waveguide}.
In such structured environments, the dispersion relation and finite bandwidth can strongly modify radiative processes and may lead to non-Markovian dynamics and bound-state formation. 
Related non-Markovian features also arise when the coupling itself is structured, for example in ``giant-atom'' waveguide QED with time-delayed self-interference \cite{kannan2020waveguide}. 
A basic and practically important question is whether one can selectively control the lifetime of one discrete degree of freedom in a multilevel system coupled to a common continuum.

Periodic driving provides an additional control knob. 
Within Floquet theory, a time-periodic Hamiltonian is described by quasienergies and drive-induced sidebands (Floquet channels) in an extended Hilbert space \cite{shirley1965solution, sambe1973steady}.
Floquet engineering can renormalize effective couplings---often with Bessel-function weights---and can open or close decay channels depending on whether Floquet sidebands overlap the environmental continuum \cite{eckardt2017colloquium, rudner2020band, dunlap1986dynamic, grossmann1991coherent}.
Most existing discussions in this context focus on controlling the dissipation of the driven degree of freedom itself.
Here we instead address whether local driving of one discrete level can activate dissipation of another, undriven level coupled to the same structured continuum.

In this work, we show that local periodic driving applied to only one discrete level can induce selective remote dissipation of another, undriven level, even when both bare discrete energies are outside the static continuum band.
We study a minimal Fano--Anderson-type model with two discrete levels coupled to a one-dimensional tight-binding continuum.
In the extended Floquet representation, periodic driving generates a ladder of drive-shifted continuum replicas, and the driven level couples to multiple Floquet channels with Bessel-function weights, enabling photon-assisted pathways that connect the undriven level to a neighboring drive-shifted continuum band (Floquet sideband).
The essential requirement for remote dissipation is that the undriven level overlaps an open Floquet sideband of the continuum that becomes accessible via drive-mediated pathways (controlled by Bessel-weighted couplings). 
In the one-dimensional tight-binding example studied here, the resulting decay can be strongly enhanced when this overlap occurs near a sideband edge due to the increased density of states; however, the mechanism itself does not rely on the band-edge singularity.
While the band-edge enhancement makes the effect especially visible in the weak-coupling regime, the underlying mechanism relies on the Floquet sideband structure and the availability of an open replica channel.
Throughout this paper, ``remote'' refers to dissipation induced on the undriven level via drive-enabled pathways mediated by the driven ladder, rather than spatial separation.

We verify these predictions by direct numerical integration of the time-dependent Schr\"odinger equation and by a complex-eigenvalue analysis of the Floquet Hamiltonian, where decay rates are extracted from the imaginary parts of resonance poles \cite{petrosky1991quantum, petrosky1997liouville, ordonez2004complex, yamada2012dynamical, kanki2017exact, yamane2017analysis, yamane2018ultrafast}.
Moreover, by tuning the drive parameters, the dominant remote pathway can be switched on and off (e.g., by exploiting zeros of Bessel functions), providing a simple route to state-selective dissipation control.
From the viewpoint of reservoir engineering, this drive-mediated remote channel provides a simple knob for state-selective reset/initialization: dissipation can be activated on a target level without directly driving it.
This may be useful when direct driving of the target state is undesirable or experimentally constrained.
The quantitative agreement between the pole-implied decay rates and the time-domain envelopes provides a direct benchmark of the pole-based description in the weak-coupling regime considered here.

\section{Model Hamiltonian}
\label{sec:model}

We consider a Fano-Anderson-type model \cite{friedrichs1948perturbation} in which multiple discrete levels are coupled to a one-dimensional tight-binding continuum (chain).
In particular, no direct coupling between discrete levels is introduced; instead, interactions arise solely through a common continuum and a local time-periodic drive.
Our aim is to examine whether dissipation of an undriven discrete level can be selectively controlled under these conditions.
Throughout this work, we set $\hbar=1$ and restrict ourselves to the single-excitation sector.

\begin{figure}
\centering
  \includegraphics[width=\linewidth]{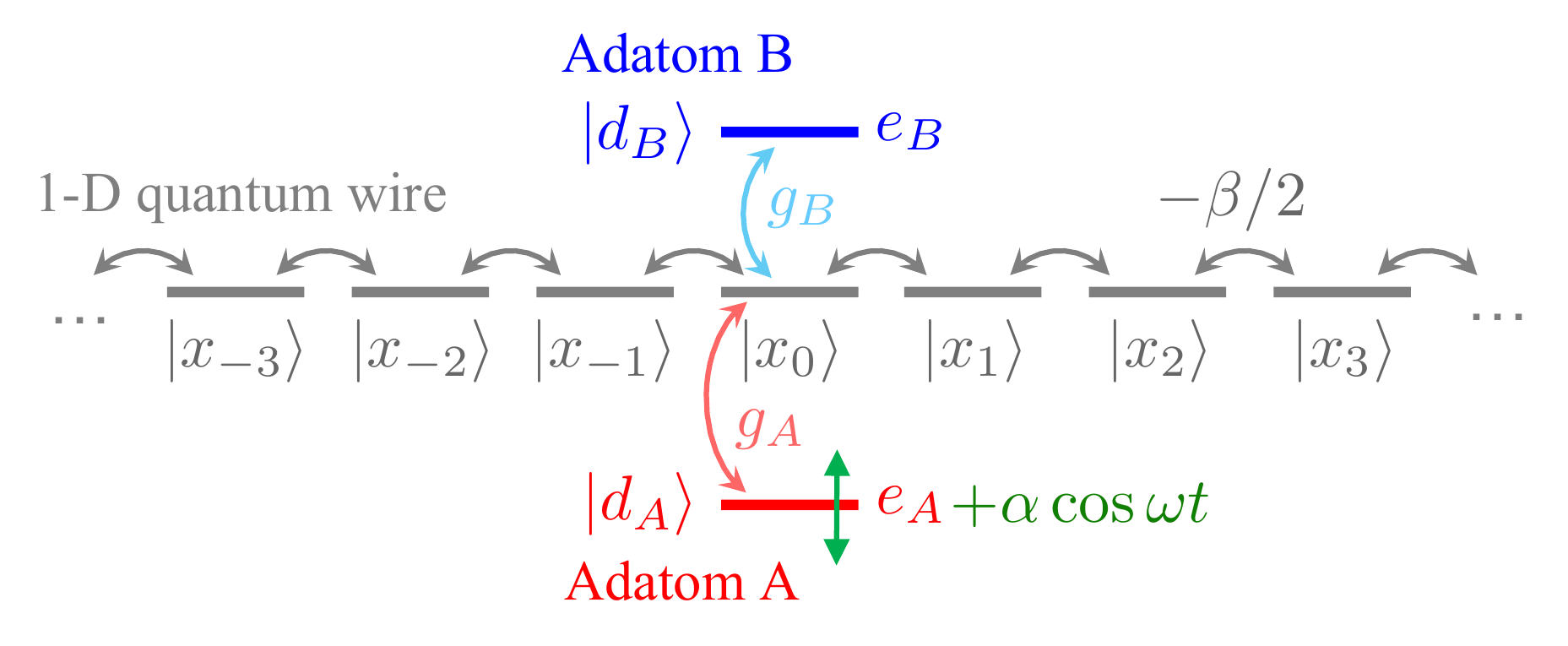}
  \caption{(Color online) Schematic picture of the model.}
  \label{fig:time}
\end{figure}

The time-dependent Hamiltonian is written as
\begin{equation}
\hat{H}(t)=\hat{H}_{\rm tb}+\hat{H}_{\rm d}(t)+\hat{H}_{\rm int},
\label{eq:H_total_braket}
\end{equation}
with
\begin{eqnarray}
\hat{H}_{\rm tb}&=&\sum_{m=-\infty}^{\infty} e_0 \ket{x_{m}}\bra{x_{m}}\nonumber\\
&&-\frac{\beta}{2}\sum_{m=-\infty}^{\infty}\left(\ketbra{x_{m+1}}{x_{m}}+\ketbra{x_{m}}{x_{m+1}}\right),\label{eq:H_chain_braket}\\
\hat{H}_{\rm d}(t)&=&\varepsilon_A(t)\ketbra{d_{A}}{d_{A}}+e_B\ketbra{d_{B}}{d_{B}},\label{eq:H_d}\\
\hat{H}_{\rm int}&=&g_A\left(\ketbra{d_{A}}{x_0}+\ketbra{x_0}{d_{A}}\right)\notag\\
&&+g_B\left(\ketbra{d_{B}}{x_0}+\ketbra{x_0}{d_{B}}\right).
\label{eq:H_int}
\end{eqnarray}
The first term describes a one-dimensional quantum chain, where $\ket{x_m}$ denotes the state localized at site $m$.
The on-site energy and nearest-neighbor hopping amplitude are given by $e_0$ and $\beta/2$, respectively.
The second term represents two discrete levels $A$ and $B$, denoted by $\ket{d_A}$ and $\ket{d_B}$, with bare energies $e_A$ and $e_B$.
Only level $A$ is subject to a time-periodic external drive,
\begin{equation}
\varepsilon_A(t)=e_A+\alpha\cos\omega t,
\end{equation}
with period $T=2\pi/\omega$.
The third term describes the coupling between the discrete levels and the chain.
The coupling constants $g_A$ and $g_B$ quantify the hybridization (hopping amplitudes) between the discrete levels and the tight-binding chain: $g_A$ ($g_B$) couples $\ket{d_A}$ ($\ket{d_B}$) to the lattice site $\ket{x_0}$.
Throughout this work we mainly focus on the weak-coupling regime,
$g_A\sim g_B\sim g \ll \beta$ (equivalently $g/\beta\ll 1$), so that the
continuum-induced decay rates are small compared with the bandwidth.
Without loss of generality, $g_A$ and $g_B$ are taken to be real.
Figure~\ref{fig:time} shows a schematic illustration of the model studied here.

The basis states are assumed to be orthonormal: $\braket{d_A}{d_A}=\braket{d_B}{d_B}=1$, $\braket{x_m}{x_{m'}}=\delta_{mm'}$, $\braket{d_{A}}{x_m}=\braket{d_{B}}{x_m}=\braket{d_A}{d_B}=0$.
No direct coupling between discrete levels is included; therefore, interactions between levels $A$ and $B$ arise solely via the continuum.

Introducing the Bloch eigenstates of the chain,
\begin{equation}
\ket{k}=\frac{1}{\sqrt{2\pi}}\sum_{m=-\infty}^{\infty} e^{ikm}\ket{x_m},
\label{eq:k_def}
\end{equation}
$k\in[-\pi,\pi)$, then we obtain
\begin{equation}
\ket{x_m}=\frac{1}{\sqrt{2\pi}}\int_{-\pi}^{\pi}dk\, e^{-ikm}\ket{k}.
\label{eq:xm_def}
\end{equation}
These states satisfy $\braket{k}{k'}=\delta(k-k')$ and
\begin{equation}
\int_{-\pi}^{\pi}dk\,\ketbra{k}{k}
=\sum_{m=-\infty}^{\infty}\ketbra{x_m}{x_m}.
\label{eq:k_orth_comp}
\end{equation}

In this basis, the Hamiltonian becomes
\begin{equation}
\hat{H}_{\rm tb}=\int_{-\pi}^{\pi}dk\,\varepsilon_k\ketbra{k}{k},
\end{equation}
with energy dispersion
\begin{equation}
\varepsilon_k=e_0-\beta\cos k,\label{eq:Htb_k}
\end{equation}
hereafter, we set the origin of energy at $e_0=0$, and the coupling between the discrete levels and the chain can be written as
\begin{eqnarray}
\hat{H}_{\rm int}&=&\frac{g_A}{\sqrt{2\pi}}\int_{-\pi}^{\pi}dk\left(\ketbra{d_A}{k}+\ketbra{k}{d_A}\right)\nonumber\\
&&+\frac{g_B}{\sqrt{2\pi}}\int_{-\pi}^{\pi}dk\left(\ketbra{d_B}{k}+\ketbra{k}{d_B}\right).\label{eq:H_int_k}
\end{eqnarray}
This completes the definition of the driven Fano-Anderson-type model studied in this work.
In the next section, we exploit the time periodicity of $\hat{H}(t)$ and formulate the problem in the extended Floquet Hilbert space.

In a straightforward generalization, the two discrete levels $A$ and $B$ may be attached to arbitrary lattice sites $\ket{x_{a}}$ and $\ket{x_{b}}$, respectively. In the Bloch representation, the couplings then acquire site-dependent phase factors $e^{-ik a}$ and $e^{-ik b}$, so that the continuum-mediated kernels depend on the separation $R\equiv|a-b|$. Although the quantitative self-energies are modified by this distance-dependent phase, the basic Floquet-assisted mechanism of remote dissipation discussed in this work remains unchanged. In the present paper, we restrict ourselves to the minimal case $a=b=0$ in order to isolate the essential physics.

\section{Floquet Formulation}
\label{sec:floquet}
To analyze the dynamics generated by the time-periodic Hamiltonian defined in Sec.~\ref{sec:model}, we employ Floquet theory, which provides a transparent energetic picture in terms of drive-shifted continuum sidebands in different Floquet channels (also called replica bands) and discrete ladders in an extended Hilbert space.
Throughout this section, $\ket{\cdot}$ denotes a state in the physical Hilbert space $\mathcal{R}$, while $\fket{\cdot}$ denotes a vector in the extended Floquet space $\mathcal{F}=\mathcal{R}\otimes\mathcal{T}$, where $\mathcal{T}$ is the space of $T$-periodic functions \cite{yamada2012dynamical, yamane2018ultrafast}. 

\begin{figure*}[t]
\centering
  \includegraphics[width=0.7\linewidth]{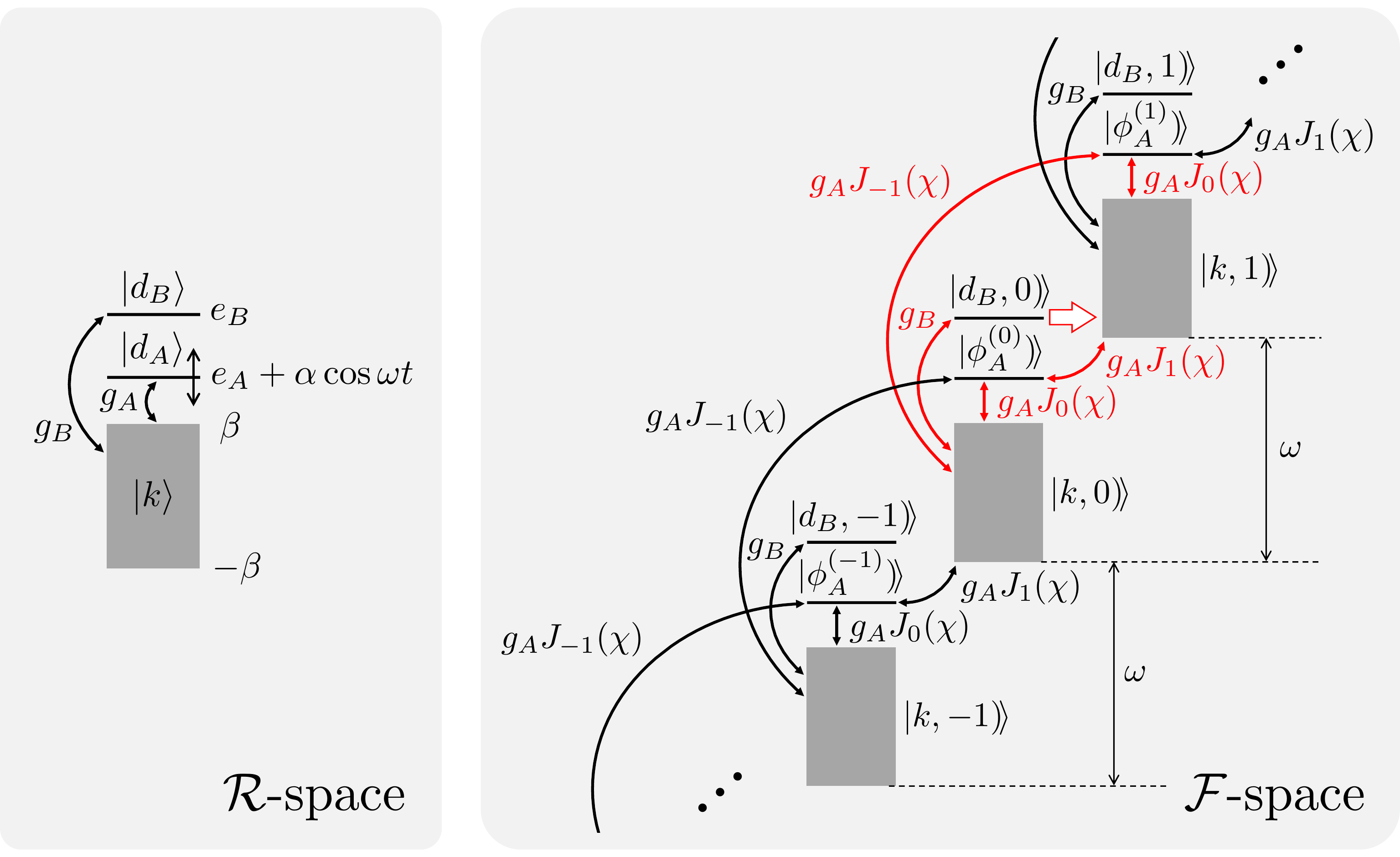}
  \caption{(Color online) Schematic energy structure in the physical ($\mathcal{R}$) and extended Floquet ($\mathcal{F}$) representations.
  The tight-binding continuum forms a finite band (gray), while the driven discrete ladder $\{\fket{\phi_A^{(n)}}\}$
  couples to different Floquet channels with Bessel-weighted amplitudes $J_{n'-n}(\chi)$.
  The undriven ladder $\{\fket{d_B,n}\}$ couples to the continuum without changing $n$.
  This structure allows photon-assisted pathways, mediated by the driven ladder, that connect the undriven state to
  a neighboring continuum replica.}
  \label{fig:energy_scheme}
\end{figure*}

Exploiting the time periodicity of the Hamiltonian, 
\begin{equation}
\hat{H}(t+T)=\hat{H}(t),\quad T=2\pi/\omega,
\end{equation}
we introduce the Floquet Hamiltonian
\begin{equation}
\hat{\mathscr{H}}\equiv \hat{H}(t)-i\frac{\partial}{\partial t}
=\sum_{n=-\infty}^{\infty}\left[\hat{\mathscr{H}}_{\rm tb}^{(n)}+\hat{\mathscr{H}}_{\rm d}^{(n)}+\hat{\mathscr{H}}_{\rm int}^{(n)}\right].
\end{equation}
In the composite Hilbert space of the physical states and time-periodic functions,
the Floquet Hamiltonian is given by
\begin{eqnarray}
\hat{\mathscr{H}}_{\rm tb}^{(n)}&=&\int_{-\pi}^{\pi}dk\,\left(\varepsilon_k+n\omega\right)\fket{k,n}\fbra{k,n},\\
\hat{\mathscr{H}}_{\rm d}^{(n)}&=&(e_A+n\omega)\fketbra{\phi_A^{(n)}}{\phi_A^{(n)}}\nonumber\\
&&+(e_B+n\omega)\fketbra{d_B,n}{d_B,n},\\
\hat{\mathscr{H}}_{\rm int}^{(n)}&=&\frac{g_A}{\sqrt{2\pi}}\int_{-\pi}^{\pi}dk\sum_{n'=-\infty}^{\infty}J_{n'-n}\left(\chi\right)\notag\\
&&\quad\quad\quad\times\left(\fketbra{\phi_A^{(n')}}{k,n}+\fketbra{k,n}{\phi_A^{(n')}}\right)\nonumber\\
&&+\frac{g_B}{\sqrt{2\pi}}\int_{-\pi}^{\pi}dk\left(\fketbra{d_B,n}{k,n}+\fketbra{k,n}{d_B,n}\right).\notag\\
\end{eqnarray}
Here we use the Fourier basis $\fket{\alpha,n}\equiv \ket{\alpha}\otimes e^{in\omega t}$ ($\alpha=k,d_A,d_B$), so that the derivative term contributes the diagonal shift $n\omega$.

Hereafter, unless otherwise stated, sums over Floquet indices are understood to run over all integers, from $-\infty$ to $\infty$, and we omit the limits for brevity.
In practical calculations, these Floquet sums are truncated only after the formal derivation.
The states $\fket{\phi_A^{(n)}}$ form the Wannier--Stark basis associated with the driven discrete level $A$, defined as
\begin{equation}
\fket{\phi_A^{(n)}}=\sum_{\nu}J_{n-\nu}\left(\chi\right)\fket{d_A,\nu}.
\end{equation}
Here $J_n(\chi)$ denotes the Bessel function of the first kind of order $n$ and $\chi\equiv\alpha/\omega$.
This Wannier--Stark basis diagonalizes the driven discrete level in the extended Floquet space,
and converts the coupling between $\fket{d_A,n}$ and $\fket{k,n}$ into Bessel-weighted couplings between
$\fket{\phi_A^{(n)}}$ and $\fket{k,n}$, which will lead directly to a Bessel-function expansion of the self-energy.

The Floquet formulation makes it transparent when dissipative channels become available.
The continuum in the $n$th Floquet channel spans the shifted band
$[-\beta+n\omega,\ \beta+n\omega]$, while the discrete ladders are located at $e_X+n\omega$ ($X=A,B$).
In the present model, the undriven ladder $\{\fket{d_B,n}\}$ is channel-diagonal in the sense that
$\fket{d_B,n}$ couples only to $\fket{k,n}$, whereas the driven ladder $\{\fket{\phi_A^{(n)}}\}$ couples
off-diagonally in $n$ through the Bessel weights.
As a result, even when the bare undriven level $e_B$ is off-resonant from the static continuum band,
it can acquire a finite linewidth via drive-induced pathways that access a neighboring continuum replica
through virtual excursions to the driven ladder.

More generally, a remote decay channel for $B$ becomes available whenever the relevant quasienergy is brought inside an open continuum replica, i.e., when $z_B-\nu\omega$ lies within the static band for some $\nu$, so that $\Im\zeta^{(\nu)}(z_B)$ becomes nonzero.
In this paper we focus on the edge-enhanced regime because it yields a particularly pronounced effect in the weak-coupling limit.

A particularly transparent situation is the first replica, where the undriven level is tuned close to the lower edge of the neighboring ($n=1$) continuum band,
\begin{equation}
e_B \simeq -\beta+\omega,
\label{eq:first_replica_resonance}
\end{equation}
while the driven level remains outside the open bands (e.g., $e_A<-\beta+\omega$ for the case shown in
Fig.~\ref{fig:energy_scheme}).
In this regime, the dominant remote-dissipation pathway for $B$ is mediated by the driven ladder and is controlled by
products of Bessel weights from neighboring Floquet channels (typically involving $J_0(\chi)$ and $J_1(\chi)$),
and its contribution to the effective self-energy enters at order $(g_A g_B)^2$.
Therefore, by appropriate choices of the energy parameters and the driving parameters $(\omega,\alpha)$, one can open
a decay channel for the undriven state $B$ through the neighboring replica of the continuum, while keeping the driven
off-resonant state $A$ long-lived.
In the next section, we verify this Floquet-energy prediction by direct numerical time evolution of the
time-dependent Schr\"odinger equation.

\section{Analytical Expression by Complex Eigenvalues}
\label{sec:complex}

In this section we derive dispersion equations for the complex quasienergies (resonance poles) of the Floquet
Hamiltonian in the extended Hilbert space $\mathcal{F}$.
The outgoing (retarded) boundary condition is imposed by the analytic continuation $z\to z^{+}$ \cite{petrosky1991quantum}.
We first eliminate the tight-binding continuum exactly and obtain an energy-dependent effective problem in the
discrete Floquet sector.
This exact formulation leads to determinant-type dispersion equations in the Floquet-index space, as in the
standard Brillouin--Wigner--Feshbach approach.
After establishing the exact dispersion condition, we then introduce a projection-operator approximation that
retains only the physical Floquet component $\fket{d_B,0}$ of the undriven ladder and obtain a scalar dispersion
equation for the $B$-like pole.

\subsection{Projection formalism and effective Floquet Hamiltonian}
\label{subsec:proj_formalism}
The resonance poles are obtained by analytic continuation of the retarded Green's function.
Here $z^{+}$ indicates the retarded prescription on the real axis, $z^{+}\equiv E+i0$,
and the resulting expressions are continued to complex $z$ to locate poles in the lower half-plane.

The resolvent of the Floquet Hamiltonian is
\begin{equation}
\hat{\mathscr{G}}(z)\equiv \frac{1}{z^{+}-\hat{\mathscr{H}}}.
\end{equation}
To eliminate the continuum in a controlled manner we introduce the projector onto the discrete sector (two ladders in Floquet space),
\begin{equation}
\hat{P}_{\rm d}\equiv\sum_{n}\left(\fketbra{\phi_A^{(n)}}{\phi_A^{(n)}}+\fketbra{d_B,n}{d_B,n}\right),
\label{eq:Pd_def}
\end{equation}
and the complementary projector onto the continuum sector,
\begin{equation}
\hat{Q}_{\rm d}\equiv \hat{I}-\hat{P}_{\rm d}=\sum_{n}\int_{-\pi}^{\pi}dk\,\fketbra{k,n}{k,n}.
\label{eq:Qd_def}
\end{equation}
We stress that the decomposition $\hat{I}=\hat{P}_{\rm d}+\hat{Q}_{\rm d}$ is not an approximation.
It is introduced solely to integrate out the continuum states $\{\fket{k,n}\}$ and to define the energy-dependent
self-energy acting on the discrete Floquet sector.  The projection approximation specific to the $B$-like pole will
be introduced later in a separate subsection.

Eliminating $\hat{Q}_{\rm d}$ by the Brillouin--Wigner--Feshbach identity yields the self-energy operator \cite{feshbach1958unified}
\begin{equation}
\hat{\Xi}(z)\equiv
\hat{P}_{\rm d}\hat{\mathscr{H}}_{\rm int}\hat{Q}_{\rm d}\,
\frac{1}{z^{+}-\hat{Q}_{\rm d}\hat{\mathscr{H}}_{\rm tb}\hat{Q}_{\rm d}}\,
\hat{Q}_{\rm d}\hat{\mathscr{H}}_{\rm int}\hat{P}_{\rm d}.
\label{eq:Xi_operator}
\end{equation}
The effective eigenvalue problem in the discrete sector reads
\begin{equation}
\left[\hat{P}_{\rm d}\hat{\mathscr{H}}_{\rm d}\hat{P}_{\rm d}+\hat{\Xi}(z)\right]\hat{P}_{\rm d}\fket{\Psi}
=z\,\hat{P}_{\rm d}\fket{\Psi}.
\label{eq:discrete_sector_eig}
\end{equation}

We define the self-energy matrices by
\begin{eqnarray}
\Xi_{AA}^{(n,n')}(z)&\equiv&\fbra{\phi_A^{(n)}}\hat{\Xi}(z)\fket{\phi_A^{(n')}},\label{eq:Xi_blocks_1}\\
\Xi_{BB}^{(n,n')}(z)&\equiv&\fbra{d_B,n}\hat{\Xi}(z)\fket{d_B,n'},\\
\Xi_{AB}^{(n,n')}(z)&\equiv&\fbra{\phi_A^{(n)}}\hat{\Xi}(z)\fket{d_B,n'},\\
\Xi_{BA}^{(n,n')}(z)&\equiv&\fbra{d_B,n}\hat{\Xi}(z)\fket{\phi_A^{(n')}}.\label{eq:Xi_blocks_2}
\end{eqnarray}
For later convenience we factor out the coupling constants and introduce coupling-independent kernels
$\xi_{XX'}^{(n,n')}(z)$:
\begin{equation}
\Xi_{XX'}^{(n,n')}(z)\equiv g_X g_{X'}\,\xi_{XX'}^{(n,n')}(z),
\qquad (X,X'=A,B).
\label{eq:Xi_factorization}
\end{equation}
\begin{align}
\xi_{AA}^{(n,n')}(z)&=\sum_{\nu}J_{n-\nu}(\chi)J_{n'-\nu}(\chi)\zeta^{(\nu)}(z),
\label{eq:xi_AA_explicit}\\
\xi_{BB}^{(n,n')}(z)&=\zeta^{(n)}(z)\delta_{nn'},\label{eq:xi_BB_explicit}\\
\xi_{AB}^{(n,n')}(z)&=J_{n-n'}(\chi)\zeta^{(n')}(z),\label{eq:xi_AB_explicit}\\
\xi_{BA}^{(n,n')}(z)&=J_{n'-n}(\chi)\zeta^{(n)}(z).\label{eq:xi_BA_explicit}
\end{align}
Here we introduced the Floquet-channel Green's function of the tight-binding chain,
\begin{equation}
\zeta^{(n)}(z)\equiv\int_{-\pi}^{\pi}\frac{dk}{2\pi}\frac{1}{z^{+}-\varepsilon_k-n\omega}=\zeta^{(0)}(z-n\omega).
\label{eq:zeta_def_complex}
\end{equation}

Using the tight-binding dispersion $\varepsilon_k=-\beta\cos k$, Eq.~\eqref{eq:zeta_def_complex}
can be recast into an integral in the energy plane.  Introducing the two-valued function
\begin{equation}
\rho(\varepsilon)\equiv\frac{1}{\sqrt{\beta^2-\varepsilon^2}},
\end{equation}
which has branch points at $\varepsilon=\pm\beta$, we take the branch cut on $\varepsilon\in[-\beta,\beta]$
and choose the values on the two rims of the cut such that
$\rho(\varepsilon+i0)=-\rho(\varepsilon-i0)=1/\sqrt{\beta^2-\varepsilon^2}$ for $\varepsilon\in(-\beta,\beta)$.
Then the Floquet-channel Green's function can be written as
\begin{equation}
\zeta^{(n)}(z)=\oint_{C}\frac{d\varepsilon}{2\pi}\frac{\rho(\varepsilon)}{z^{+}-\varepsilon-n\omega},
\label{eq:zeta_energy_integral}
\end{equation}
Here $C$ represents a contour encircling the branch cut associated with the continuum spectrum.
The prescription $z^{+}$ indicates that the resolvent is defined by analytic continuation of $z$
from the upper half of the complex plane, corresponding to the retarded boundary condition.
Under this analytic continuation, the integration contour $C$ is implicitly deformed so as to remain
consistent with the chosen Riemann sheet of $\zeta^{(n)}(z)$.
In this sense, the contour integral representation incorporates both the branch-cut structure of the
continuum and the analytic continuation required to describe resonance poles in the complex energy plane.
Such a deformation of integration contours under analytic continuation is a standard procedure in the
complex spectral representation of open quantum systems
\cite{petrosky1991quantum}.

This structure also clarifies how photon-assisted processes can open decay channels.
Even if a bare discrete energy is off-resonant from the static ($n=0$) continuum band, exchange of $n$ drive
quanta can bring it into resonance with a continuum replica shifted by $n\omega$.
In the present model, such access to neighboring replicas is enabled by the driven ladder through the
Bessel-weighted off-diagonal Floquet couplings, and it will be responsible for the selective remote dissipation
discussed below.
At this stage, we keep the integral representation and use only its analytic properties, without evaluating it further.

\subsection{Exact dispersion equations in the discrete Floquet sector}
\label{subsec:disp_exact}

We now derive the exact dispersion conditions for the complex quasienergies by projecting
Eq.~\eqref{eq:discrete_sector_eig} onto the two discrete ladders in Floquet space.
The discrete-sector component of a Floquet eigenvector can be expanded as
\begin{equation}
\hat{P}_{\rm d}\fket{\Psi}
=\sum_{n}\left(u_{A}^{(n)}\fket{\phi_A^{(n)}}+u_{B}^{(n)}\fket{d_B,n}\right),
\label{eq:PdPsi_expansion}
\end{equation}
where
\begin{equation}
u_{A}^{(n)}\equiv \fbraket{\phi_A^{(n)}}{\Psi},
\qquad
u_{B}^{(n)}\equiv \fbraket{d_B,n}{\Psi}.
\end{equation}
Projecting Eq.~\eqref{eq:discrete_sector_eig} onto the ladder bases yields the coupled homogeneous equations
for $X,Y=A,B$ and $X\neq Y$,
\begin{align}
\sum_{n'}\Delta_{X}^{(n,n')}(z)\,u_{X}^{(n')}=g_{X}g_{Y}\sum_{n'}\xi_{XY}^{(n,n')}(z)\,u_{Y}^{(n')},
\label{eq:discrete_eq}
\end{align}
where we introduced the inverse Green's-function kernels
\begin{align}
\Delta_{X}^{(n,n')}(z)
&\equiv
\left(z-e_X-n\omega\right)\delta_{nn'}-g_{X}^{2}\xi_{XX}^{(n,n')}(z).
\label{eq:Delta_def}
\end{align}

For notational symmetry we define the corresponding Green matrices
\begin{equation}
{\bm G}_X(z)\equiv \left[{\bm \Delta}_X(z)\right]^{-1}.
\label{eq:GA_GB_def}
\end{equation}
Then Eq.~\eqref{eq:discrete_eq} can be written as the formal solutions
\begin{align}
u_{X}^{(n)}&=g_{X}g_{Y}\sum_{\mu,\mu'}G_X^{(n,\mu)}(z)\xi_{XY}^{(\mu,\mu')}(z)u_{Y}^{(\mu')}.
\label{eq:u_formal_symmetric}
\end{align}
Eliminating $\{u_{Y}^{(n)}\}$ from Eq.~\eqref{eq:discrete_eq} by using Eq.~\eqref{eq:u_formal_symmetric},
we obtain a closed homogeneous system for $\{u_{X}^{(n)}\}$,
\begin{equation}
\sum_{n'}\eta_X^{(n,n')}(z)\,u_{X}^{(n')}=0,
\label{eq:etaA_linear}
\end{equation}
with
\begin{eqnarray}
\eta_X^{(n,n')}(z)
&\equiv&
(z-e_X-n\omega)\delta_{nn'}
-g_{X}^{2}\xi_{XX}^{(n,n')}(z)\nonumber\\
&&-g_{X}^{2}g_{Y}^{2}\sum_{\mu,\mu'}\xi_{XY}^{(n,\mu)}(z)\,G_Y^{(\mu,\mu')}(z)\,\xi_{YX}^{(\mu',n')}(z).\nonumber\\
\label{eq:etaA_matrix_symmetric}
\end{eqnarray}
A nontrivial solution $\{u_{X}^{(n)}\}\neq 0$ exists if and only if
\begin{equation}
\eta_X(z)\equiv
\det_{n,n'\in\mathbb{Z}}\left[\eta_X^{(n,n')}(z)\right]=0.
\label{eq:etaA_det_symmetric}
\end{equation}
Equation~\eqref{eq:etaA_det_symmetric} gives the exact $X(=A,B)$-like complex quasienergies.
In practice, the Floquet indices are truncated only after the formal derivation.
This is an infinite-dimensional determinant in Floquet space.

\subsection{Projection onto $\fket{\phi_A^{(0)}}$ and scalar dispersion equation}
\label{subsec:disp_projected_A0}

Having established the exact determinant condition \eqref{eq:etaA_det_symmetric}, we now derive a practical
scalar dispersion equation for the $A$-like pole.
For the driven ladder $A$, photon-assisted channels are accessed directly through the Bessel-weighted couplings,
so an $A$-like quasi-energy can acquire a finite imaginary part already at ${\cal O}(g_A^2)$ whenever a shifted
channel overlaps a continuum replica, as encoded in the analytic structure of $\zeta^{(\nu)}(z)$
[Eq.~\eqref{eq:zeta_energy_integral}].

Accordingly, for the purpose of identifying the leading resonance mechanism of the driven ladder, we truncate
the kernel $\eta_A^{(n,n')}(z)$ at ${\cal O}(g_{A}^{2})$ by neglecting the inter-ladder contribution proportional to
$g_A^2 g_B^2$ in Eq.~\eqref{eq:etaA_matrix_symmetric}.

To obtain a scalar dispersion function, we employ a channel-diagonal approximation for the driven-ladder self-energy,
\begin{equation}
\xi_{AA}^{(n,n')}(z)\simeq \xi_{AA}^{(n)}(z)\delta_{nn'},
\label{eq:xi_diag_approx}
\end{equation}
with
\begin{equation}
\xi_{AA}^{(n)}(z)=\sum_{\nu}J_{n-\nu}^{2}(\chi)\zeta^{(\nu)}(z).
\label{eq:XiAA_diag}
\end{equation}
Under Eq.~\eqref{eq:xi_diag_approx}, each Wannier--Stark rung decouples and yields a scalar inverse propagator
\begin{equation}
\widetilde{\eta}_A^{(n)}(z)\equiv
\Delta_A^{(n)}(z)
\equiv z-e_A-n\omega-g_A^2\xi_{AA}^{(n)}(z).
\label{eq:etaA_tilde_def}
\end{equation}
In particular, the $A$-like pole continuously connected to the bare quasi-energy $e_A$ (modulo $\omega$) is obtained
from the $n=0$ equation $\widetilde{\eta}_A^{(0)}(z)=0$,
\begin{equation}
\widetilde{\eta}_A^{(0)}(z)
= z-e_A-g_A^2\sum_{\nu}J_{\nu}^{2}(\chi)\zeta^{(\nu)}(z),
\label{eq:etaA_tilde_0}
\end{equation}
where we used $J_{-\nu}^{2}(\chi)=J_{\nu}^{2}(\chi)$.

We emphasize that, unlike the undriven level $B$ placed outside the static band, the driven ladder can access
$\zeta^{(\nu\neq 0)}(z)$ directly through the Bessel weights, and therefore an $A$-like linewidth may emerge already at
${\cal O}(g_A^2)$ when a photon-shifted channel overlaps a continuum replica.
This contrasts with the remote-dissipation mechanism for $B$, where the first nonzero imaginary part of the $B$-like pole
arises only at higher order in the couplings, motivating the deeper expansion of the projected scalar function
$\widetilde{\eta}_B^{(0)}(z)$ in the next subsection.

\subsection{Projection onto $\fket{d_B,0}$ and scalar dispersion equation}
\label{subsec:disp_projected_B0}
We now introduce the projection-operator approximation adopted in this work: we retain only the physical Floquet
component $\fket{d_B,0}$ of the undriven ladder,
\begin{equation}
u_{B}^{(n)}\simeq \delta_{n0}\,u_B,
\qquad
u_B\equiv u_B^{(0)}.
\label{eq:proj_B0}
\end{equation}
This approximation is designed to extract the $B$-like pole continuously connected to the bare energy $e_B$ in the
weak-coupling limit.  Since the undriven ladder is channel-diagonal exactly
[Eq.~\eqref{eq:xi_BB_explicit}], the discarded $B$-replica components are generated only through inter-ladder
excursions mediated by $A$, and their feedback to the $n=0$ pole enters beyond the leading inter-ladder order.

In the following we adopt the channel-diagonal approximation \eqref{eq:xi_diag_approx} introduced above, so that the
driven-ladder propagator is given by $\Delta_A^{(n)}(z)$ [Eq.~\eqref{eq:etaA_tilde_def}].
This approximation neglects continuum-induced coherences between different Wannier--Stark rungs of the driven ladder and
is expected to be accurate in the weak-coupling regime when the characteristic linewidth is small compared with the
Floquet spacing $\omega$ and the relevant Bessel weights are concentrated in a limited number of channels.
Under Eq.~\eqref{eq:xi_diag_approx}, the driven-ladder propagator becomes channel-diagonal with $\Delta_A^{(n)}(z)$.

Substituting the explicit inter-ladder kernels
[Eqs.~\eqref{eq:xi_AB_explicit} and \eqref{eq:xi_BA_explicit}] together with
$\Delta_B^{(0,0)}(z)=z-e_B-g_B^2\zeta^{(0)}(z)$, we obtain the scalar dispersion function
\begin{eqnarray}
\widetilde{\eta}_{B}^{(0)}(z)&\equiv&z-e_B-g_B^2\zeta^{(0)}(z)\notag\\
&&-g_{A}^{2}g_{B}^{2}\left[\zeta^{(0)}(z)\right]^2\sum_{n}\frac{J_{n}^{2}(\chi)}{\Delta_A^{(n)}(z)}.
\label{eq:disp_B0_scalar_exact}
\end{eqnarray}
The projected $B$-like complex quasi-energy is determined by
$\widetilde{\eta}_{B}^{(0)}(z)=0$.

To make the coupling-order structure transparent, we further expand the driven-ladder propagator
$\{ \Delta_A^{(n)}(z)\}^{-1}$ in a Maclaurin series with respect to $g^2$.
In the weak-coupling regime $g\ll\beta$ this yields an asymptotic expansion of
$\widetilde{\eta}_{B}^{(0)}(z)$ in powers of $g$.
Keeping terms up to ${\cal O}(g_A^{4}g_B^{2})$ (and discarding ${\cal O}(g_A^{6}g_B^{2})$ and higher),
we obtain
\begin{eqnarray}
\widetilde{\eta}_{B}^{(0)}(z)&\simeq&z-e_B-g_B^2\zeta^{(0)}(z)\notag\\
&&-g_{A}^{2}g_{B}^{2}\left[\zeta^{(0)}(z)\right]^2\sum_{n}\frac{J_{n}^{2}(\chi)}{z-e_{A}-n\omega}\notag\\
&&-g_{A}^{4}g_{B}^{2}\left[\zeta^{(0)}(z)\right]^2\sum_{n,\nu}\frac{J_{n}^{2}(\chi)J_{n-\nu}^{2}(\chi)}{(z-e_{A}-n\omega)^2}\zeta^{(\nu)}(z).\notag\\
\label{eq:disp_B0_scalar}
\end{eqnarray}
In practical calculations we truncate the Floquet sums after the formal derivation and select the
$B$-like solution continuously connected to $e_B$ as $g_A,g_B\to 0$.

To summarize, starting from the exact determinant formulation in the discrete Floquet sector, we introduced two
controlled simplifications tailored to the $B$-like pole: (i) the projection \eqref{eq:proj_B0} onto the physical
component $\fket{d_B,0}$ of the undriven ladder, and (ii) the channel-diagonal approximation
\eqref{eq:xi_diag_approx} for the driven-ladder self-energy.
These steps reduce the infinite-dimensional determinant problem to the scalar dispersion equation
\eqref{eq:disp_B0_scalar_exact}.
Furthermore, the weak-coupling expansion \eqref{eq:disp_B0_scalar} exposes the coupling-order hierarchy responsible
for remote dissipation, clarifying why the $B$-like linewidth can emerge only at high order through
photon-assisted pathways that access $\zeta^{(\nu\neq 0)}(z)$.
In the following sections we solve the truncated scalar equation and compare the resulting pole-implied decay rate
with direct time-domain simulations.

\section{Selective Dissipation Induced by Local Driving}
\label{sec:selective}

In this section we demonstrate, in the time domain, that a local periodic drive applied only to the discrete level
$A$ can induce dissipation of the other, undriven level $B$ in a highly selective manner.
The numerical results are compared with the complex-quasi-energy (resonance-pole) analysis developed in
Sec.~\ref{sec:complex}.

\subsection{Time evolution and survival probabilities}
\label{subsec:time_domain_def}

We characterize the time-domain dynamics by the time-evolution operator of the periodically driven Hamiltonian
$\hat H(t)$,
\begin{equation}
\ket{\psi(t)}=\hat U(t,0)\ket{\psi(0)},
\end{equation}
\begin{equation}
\hat U(t,0)=\mathcal{T}\exp\left[-i\int_{0}^{t}d\tau\,\hat H(\tau)\right],
\label{eq:U_def}
\end{equation}
where $\mathcal{T}$ denotes time ordering.
We evaluate the survival probabilities of the two discrete states by performing two independent time evolutions
under the same Hamiltonian parameters:
for the initial preparation $\ket{\psi(0)}=\ket{d_X}$ ($X=A,B$), we define
\begin{equation}
P_X(t)\equiv|\bra{d_X}\hat U(t,0)\ket{d_X}|^2.
\label{eq:PX_def}
\end{equation}
The numerical evaluation of $P_X(t)$ is carried out by directly integrating the time-dependent
Schr\"odinger equation in the single-excitation sector. The implementation details (finite chain truncation,
absorbing boundary, time stepping, and convergence checks) are summarized in Appendix.
Here we focus on the physical interpretation and on the comparison with the pole prediction.

For later reference, we denote the complex quasi-energy (pole) of the $X$-like resonance by
\begin{equation}
z_X=\bar{e}_X-i\gamma_X,
\qquad \gamma_X\ge 0,
\end{equation}
so that the pole predicts an exponential envelope $P_X(t)\propto e^{-2\gamma_X t}$ in an intermediate-time window
where a single pole dominates.

\subsection{Decay of $B$ near the first-replica edge and validation of the pole prediction}
\label{subsec:Bexp_validation}

\begin{figure}[t]
  \centering
  \includegraphics[width=\linewidth]{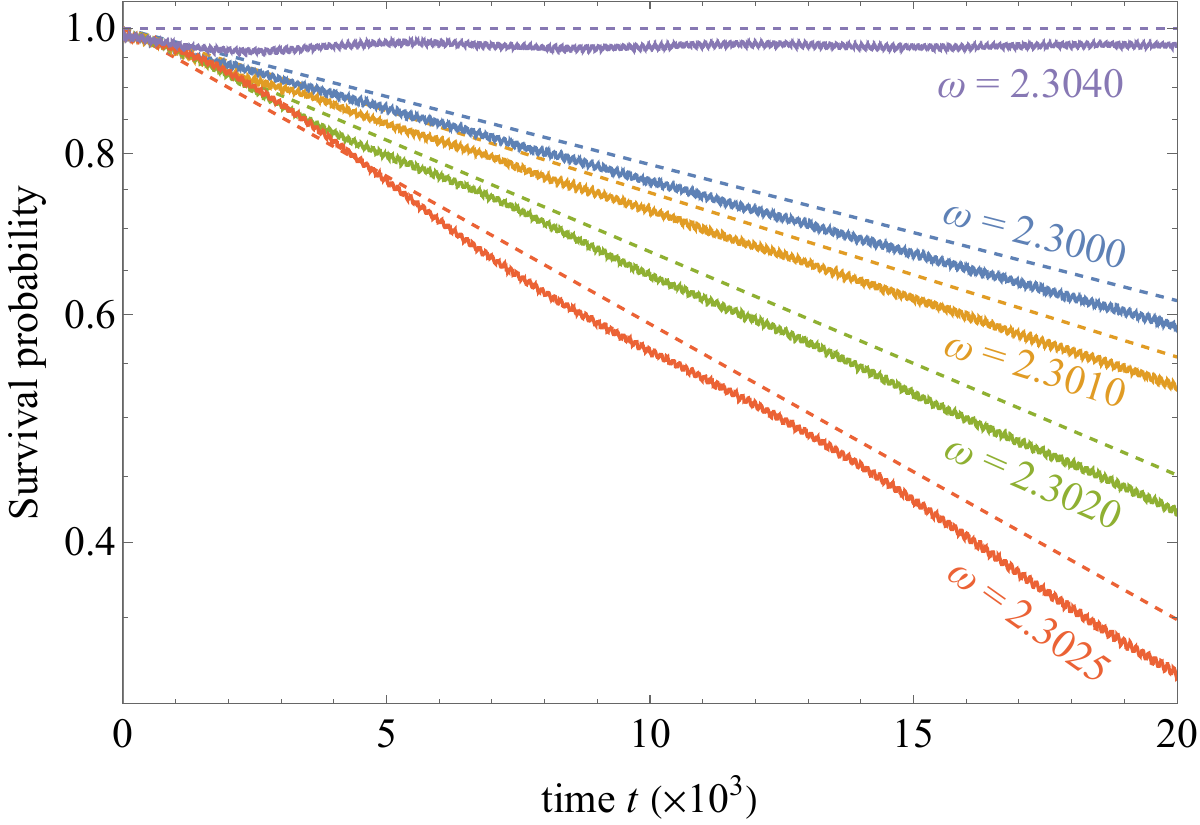}
  \caption{(Color online) Time-domain decay of the undriven state $B$ near the first-replica edge and comparison with the pole prediction. Solid curves: numerical evaluation of $P_B(t)$, with $\chi=1.081978$ and $e_0=0$, $\beta=1$, $e_A=1.25$, $e_B=1.30$, $g_A=g_B=0.05$. Dashed lines: exponential envelopes $e^{-2\gamma_{B}t}$ using $\gamma_{B}=-\Im z_B$ obtained from the complex pole.}
  \label{fig:Bexp}
\end{figure}

\begin{figure*}[t]
  \centering
  \includegraphics[width=0.48\linewidth]{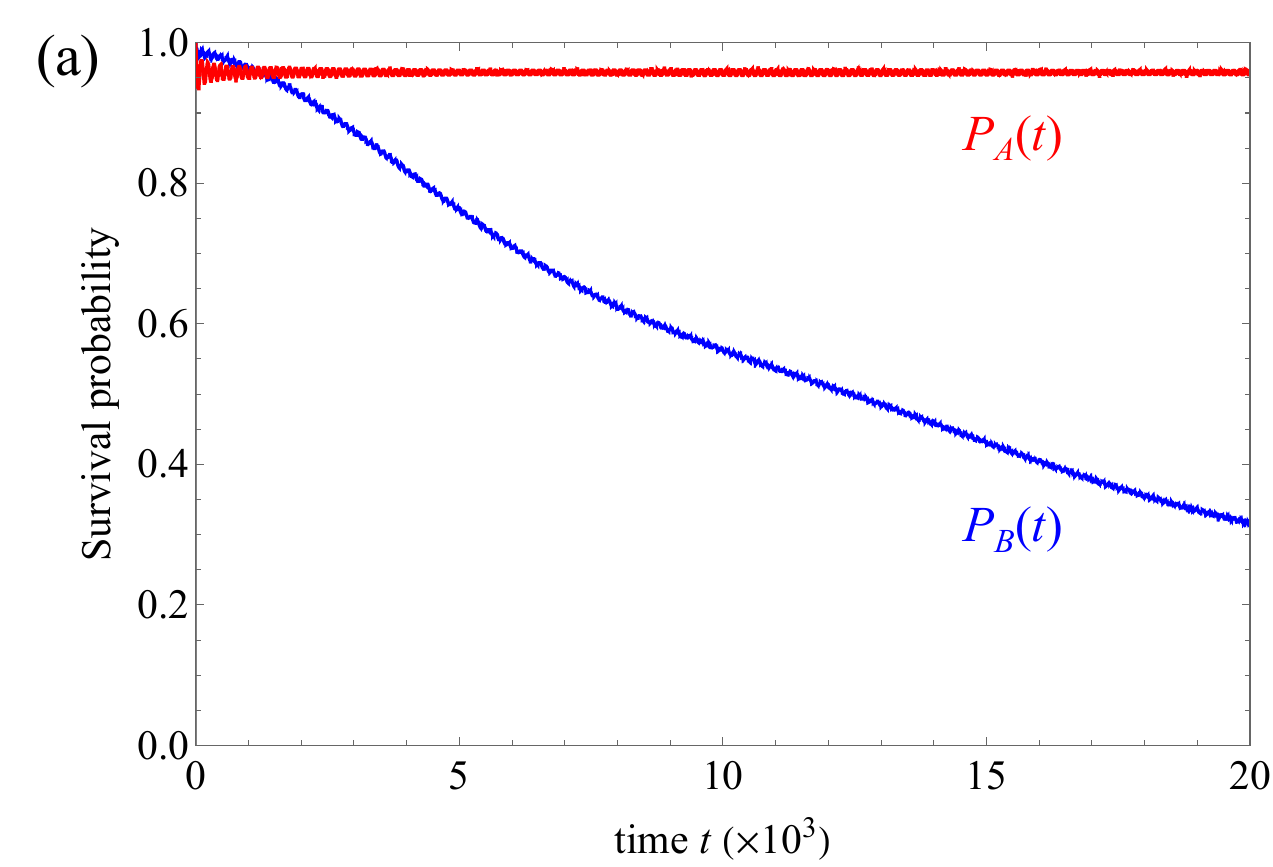}
  \hfill
  \includegraphics[width=0.48\linewidth]{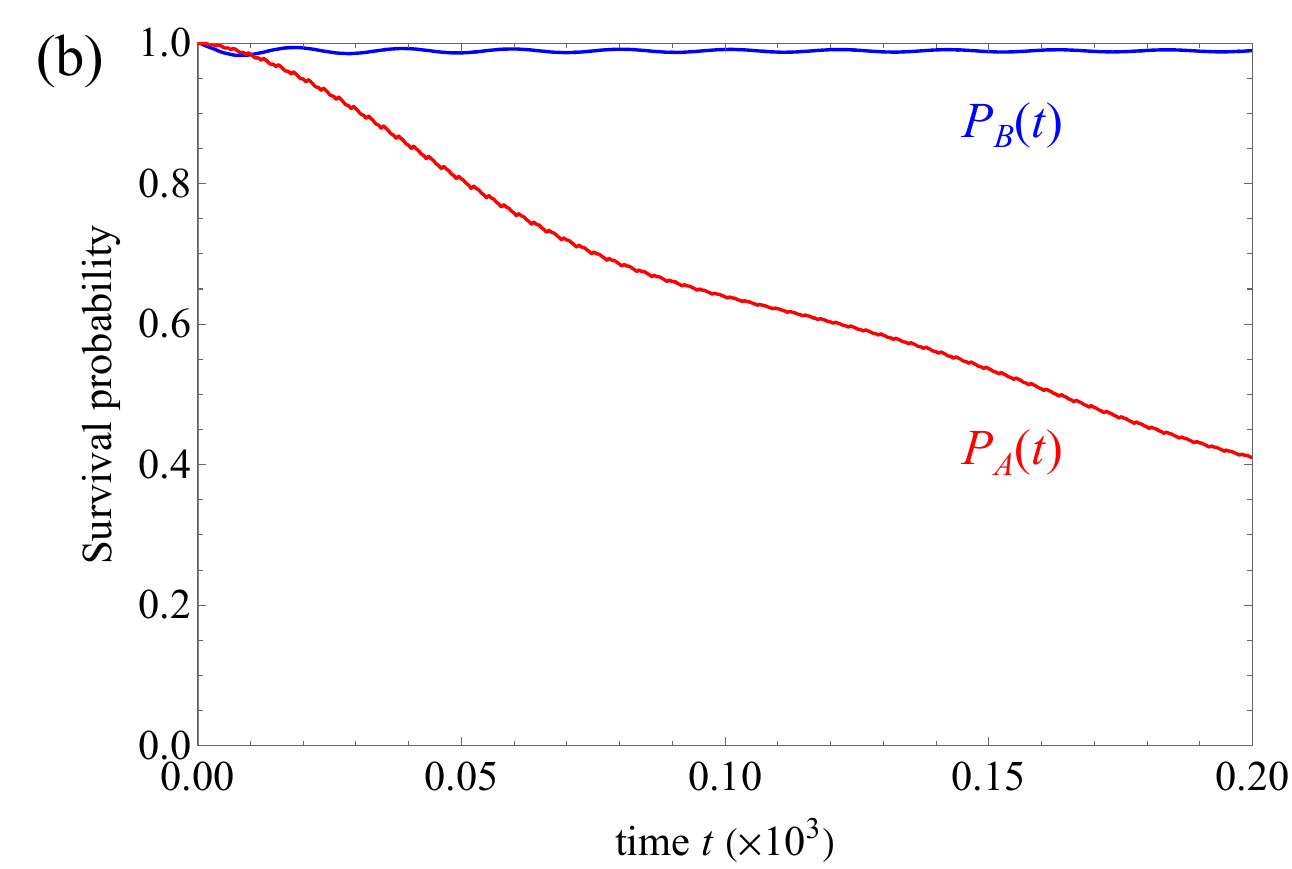}
  \caption{(Color online) Selective dissipation controlled by local driving.
  (a) Remote dissipation of the undriven level: $P_A(t)$ (red) remains close to unity while $P_B(t)$ (blue) decays for
  $\omega=2.3025$ and $\chi=1.081978$.
  (b) Switching and suppression of the remote channel: by choosing $\chi$ at the first zero of $J_0$, the decay of $B$
  is strongly suppressed while the driven level $A$ decays for $\omega=2.2$ and $\chi=2.404826$.
  The internal parameters are $e_0=0$, $\beta=1$, $e_A=1.25$, $e_B=1.30$, and $g_A=g_B=0.05$.}
  \label{fig:selective_examples}
\end{figure*}

We first demonstrate that the decay of the undriven level $B$ observed in the time domain is quantitatively
consistent with the decay constant extracted from the $B$-like resonance pole.
We fix the internal parameters to $e_0=0$, $\beta=1$, $e_A=1.25$, $e_B=1.30$, $g_A=g_B\equiv g=0.05$, and sweep the drive frequency $\omega$ in a narrow window where $e_B$ is brought into the $n=1$ continuum replica.
We focus on the vicinity of its lower edge to maximize the effect in the weak-coupling regime.

Since the tight-binding dispersion $\varepsilon_k=e_0-\beta\cos k$ yields a band $[\,-\beta,+\beta\,]$ in each Floquet
channel, the $n$th replica spans the interval
\begin{equation}
E_{\rm L}^{(n)}=e_0-\beta+n\omega,\qquad E_{\rm U}^{(n)}=e_0+\beta+n\omega.
\end{equation}
Thus, the ``first-replica edge'' for the lower boundary corresponds to $E_{\rm L}^{(1)}=-\beta+\omega$, and the
condition $e_B\simeq E_{\rm L}^{(1)}$ becomes $\omega\simeq e_B+\beta\simeq 2.3$.

Throughout this section we parameterize the drive by $\chi\equiv\alpha/\omega$ and choose $\chi=1.081978$, which maximizes $|J_0(\chi)J_1(\chi)|$ and therefore enhances the leading Floquet-assisted remote channel responsible
for the decay of $B$ in the present configuration.

This parameter choice also allows us to identify the leading microscopic origin of the linewidth of the $B$-like pole.
In the present configuration, $e_B$ lies outside the static ($n=0$) band, so the direct term
$-g_B^2\zeta^{(0)}(z)$ in the dispersion function is real and does not generate decay.
Moreover, the ${\cal O}(g_A^{2}g_B^{2})$ contribution remains predominantly real here because it involves only
$\zeta^{(0)}(z)$ evaluated at $z\simeq e_B\notin[-\beta,\beta]$.
The first contribution that can generate a nonzero imaginary part is the ${\cal O}(g_A^{4}g_B^{2})$ term in
Eq.~\eqref{eq:disp_B0_scalar}, which contains $\zeta^{(\nu)}(z)$.
Near the first-replica-edge condition $e_B\simeq -\beta+\omega$, the $\nu=1$ component
$\zeta^{(1)}(z)=\zeta^{(0)}(z-\omega)$ is evaluated close to the static band edge, producing a large negative
imaginary part and hence the linewidth of the $B$-like pole.

Furthermore, the sum over $n$ in Eq.~\eqref{eq:disp_B0_scalar} is dominated by $n=0$ for our parameter set
because $|z-e_A|\ll\omega$ while $|z-e_A-n\omega|\sim{\cal O}(\omega)$ for $n\neq 0$.
As a result, the leading drive dependence of the linewidth is governed by the neighboring-channel Bessel weights,
\begin{equation}
\gamma_B \propto g_A^{4}g_B^{2}\,J_0^{2}(\chi)J_1^{2}(\chi),
\label{eq:gammaB_scaling}
\end{equation}
up to slowly varying factors associated with detunings and the local Green's functions.
This scaling is consistent with the intuitive Floquet quasi-energy picture of the remote pathway mediated by the
neighboring continuum replica.

Figure~\ref{fig:Bexp} shows $P_B(t)$ obtained from the numerical time evolution (solid curves) for the representative
drive frequencies $\omega=2.3000$, $2.3010$, $2.3020$, $2.3025$, and $2.3040$.
For each $\omega$, we solve the dispersion equation in Sec.~\ref{sec:complex} and obtain the $B$-like pole
$z_B\equiv\bar{e}_B-i\gamma_{B}$, which yields
$\gamma_B\simeq 1.2133\times10^{-5}$,
$1.4645\times10^{-5}$,
$1.9906\times10^{-5}$,
$2.6346\times10^{-5}$,
and $0$ for the above five frequencies in this order.
Here $\gamma_B\simeq 0$ indicates that the corresponding pole is purely real within numerical resolution.
The dashed lines in Fig.~\ref{fig:Bexp} plot the exponential envelopes $e^{-2\gamma_{B}t}$ using these pole-implied
decay constants.
The agreement confirms that the complex-quasi-energy description captures the time-domain decay of the undriven level in the regime where it overlaps the neighboring continuum replica; in the present tight-binding setting the decay is particularly pronounced near the replica edge due to the enhanced density of states.
Moreover, the vanishing linewidth at $\omega=2.3040$ is consistent with the interpretation that, for a sufficiently
large drive frequency, $e_B$ no longer resonates with the neighboring ($n=1$) continuum replica and the
Floquet-assisted decay channel closes.

\subsection{Remote dissipation of undriven $B$ while keeping driven $A$ long-lived}
\label{subsec:remote_B_only}

We now highlight the central and counterintuitive consequence of local driving:
although only $A$ is driven, the undriven level $B$ can be made to decay while the driven level $A$ remains long-lived.
This regime is realized when $e_B$ is placed inside (or close to the edge of) a neighboring replica, $e_B\simeq E_{\rm L}^{(1)}$, so that a strong dissipative channel is available for $B$ through drive-enabled pathways mediated by the driven ladder, whereas $A$ remains effectively off-resonant from all open replica intervals with appreciable Bessel weight.

Figure~\ref{fig:selective_examples}(a) shows the survival probabilities $P_A(t)$ (red) and $P_B(t)$ (blue) for
$\omega=2.3025$ and $\chi=1.081978$ with the same internal parameters as in
Sec.~\ref{subsec:Bexp_validation}.
Despite the drive acting on $A$, the $A$-survival probability stays close to unity over the shown time window,
indicating that the driven level is effectively stabilized in this off-resonant configuration.
In contrast, $P_B(t)$ exhibits a pronounced decay, demonstrating selective remote dissipation of the undriven level.

\subsection{Switching the dissipation channel by tuning $\chi$: suppressing the remote decay of $B$}
\label{subsec:suppress_B_by_J0}

Finally, we demonstrate that the selective dissipation pathway can be switched by tuning only the external driving
parameters $(\omega,\alpha)$, without changing the internal system parameters.
We choose a configuration where the driven level has an energetically allowed Floquet-assisted decay channel
(e.g., $e_A$ lies inside the $n=1$ replica interval), so that $A$ can decay at the leading order
$\Gamma_A\sim \mathcal{O}(g_A^2)$.
By contrast, the decay of $B$ remains remote and is controlled by the Floquet-assisted pathway weighted by
$J_0(\chi)J_1(\chi)$.
Therefore, choosing $\chi$ at a zero of $J_0(\chi)$ suppresses the dominant remote channel of $B$ while leaving the
$A$-decay channel open as long as $J_1(\chi)\neq 0$.
In other words, even though both $A$ and $B$ are resonant with replica bands and thus unstable, it is possible to engineer a regime where only $B$ exhibits dynamical localization.

Figure~\ref{fig:selective_examples}(b) shows $P_A(t)$ (red) and $P_B(t)$ (blue) for $\omega=2.2$ and $\chi=2.404826$ (i.e., $J_0(\chi)=0$), with the same internal parameters as above.
In this case $P_A(t)$ decays rapidly, consistent with the availability of an open replica channel for the driven level,
whereas $P_B(t)$ remains close to unity: the undriven level is stabilized by closing the leading remote pathway through
the Bessel-function control knob.

\section{Conclusion}
\label{sec:conclusion}
We studied a minimal driven Fano-Anderson-type model in which two discrete levels are coupled to a one-dimensional tight-binding continuum, while a local periodic drive is applied only to one of the levels. 
We addressed whether dissipation of an undriven discrete state can be activated selectively and remotely, even when both bare discrete energies are off-resonant from the static ($n=0$) continuum band.

In the extended Floquet representation, the continuum forms a ladder of shifted finite bands, whereas the driven ladder couples to multiple Floquet channels with Bessel-function weights. 
This energetic structure implies that the undriven level can acquire a finite linewidth via drive-enabled pathways mediated by virtual excursions to the driven ladder. 
In particular, when the undriven level is brought to overlap an open continuum replica, it can acquire a finite linewidth even though it lies outside the static band; in the present tight-binding example the effect is especially pronounced near a replica-band edge (replica-edge condition) where the density of states is enhanced.

We verified these predictions by direct numerical time evolution for two independent initial preparations and observed a pronounced decay of the undriven state in the replica-edge regime, whereas the driven off-resonant state remains long-lived over the same time window.

To quantify the decay rates and elucidate the mechanism, we formulated the complex-eigenvalue (resonance-pole) problem of the Floquet Hamiltonian by eliminating the continuum via a Brillouin--Wigner--Feshbach projection. Within the weak-coupling regime and a channel-diagonal approximation for the intra-ladder self-energy, the imaginary parts of the relevant complex quasienergies reproduce the time-domain decay envelopes. In the remote-dissipation regime, the undriven (B-like) pole acquires a finite negative imaginary part when the relevant quasienergy overlaps an open replica band, with a pronounced enhancement in the vicinity of the replica-edge condition, whereas the driven (A-like) pole remains nearly real.

Importantly, the remote dissipation pathway can be switched and suppressed by tuning only the driving parameters: for the first replica, the leading remote channel is controlled by Bessel-weight products (typically involving $J_{0}(\chi)$ and $J_{1}(\chi)$), so choosing $\chi$ near a zero of $J_{0}(\chi)$ strongly suppresses the remote decay of the undriven level.

Our results provide a simple route to state-selective dissipation control using local periodic driving in structured environments, without requiring direct coupling between discrete levels.

\begin{acknowledgment}
This work was supported by JSPS KAKENHI Grant Number JP24K00933.
\end{acknowledgment}

\appendix
\section{Numerical Simulation of the Time Evolution}
\label{sec:time_evolution}
In this Appendix we summarize the numerical procedure used to obtain the time-domain survival probabilities
shown in Sec.~\ref{sec:selective}.
Rather than constructing the time-evolution operator $\hat U(t,0)$ explicitly, we directly integrate the
time-dependent Schr\"odinger equation under the Hamiltonian \eqref{eq:H_total_braket} and evaluate
$P_X(t)=|\bra{d_X}\hat U(t,0)\ket{d_X}|^2$ from the discrete-state amplitudes.

The time evolution is defined by
\begin{equation}
\ket{\psi(t)}=\hat U(t,0)\ket{\psi(0)},
\end{equation}
\begin{equation}
\hat U(t,0)=\mathcal{T}\exp\left[-i\int_{0}^{t}d\tau\,\hat H(\tau)\right],
\label{eq:U_def_app}
\end{equation}
where $\mathcal{T}$ denotes time ordering.
For the initial preparation $\ket{\psi(0)}=\ket{d_X}$ ($X=A,B$), we evaluate the survival probability
\begin{equation}
P_X(t)\equiv|\bra{d_X}\hat U(t,0)\ket{d_X}|^2.
\label{eq:PX_def_U}
\end{equation}
In the numerical integration described below, this quantity is simply given by
$P_A(t)=|d_A(t)|^2$ and $P_B(t)=|d_B(t)|^2$, because $\ket{\psi(0)}=\ket{d_X}$ implies
$\bra{d_X}\psi(t)\rangle=d_X(t)$.

We solve
\begin{equation}
i\frac{d}{dt}\ket{\psi(t)}=\hat{H}(t)\ket{\psi(t)}.
\label{eq:TDSE}
\end{equation}
In the single-excitation sector, we expand the state vector as
\begin{equation}
\ket{\psi(t)}=
d_{A}(t)\ket{d_A}+d_{B}(t)\ket{d_B}+\sum_{m=-M}^{M}c_m(t)\ket{x_m},
\label{eq:psi_expand}
\end{equation}
where the tight-binding chain is truncated to a finite interval $m=-M,\ldots,M$.
Substituting Eq.~\eqref{eq:psi_expand} into Eq.~\eqref{eq:TDSE} yields the coupled equations of motion
\begin{eqnarray}
i\dot{d}_{A}(t) &=& \varepsilon_A(t)\,d_{A}(t)+g_A\,c_{0}(t),
\label{eq:eom_a}\\
i\dot{d}_{B}(t) &=& e_B\,d_{B}(t)+g_B\,c_{0}(t),
\label{eq:eom_b}\\
i\dot{c}_{m}(t) &=& e_0\,c_{m}(t)-\frac{\beta}{2}\left[c_{m+1}(t)+c_{m-1}(t)\right]\notag\\
&&+g_A\,d_{A}(t)\delta_{m,0}+g_{B}\,d_{B}(t)\delta_{m,0}.
\label{eq:eom_cm}
\end{eqnarray}

To suppress unphysical reflections from the finite boundaries at $m=\pm M$, we employ a complex absorbing
potential (CAP) implemented as an imaginary on-site term on the edge sites~\cite{muga2004complex},
\begin{equation}
\hat{H}_{\rm CAP}=-i\sum_{m=-M}^{M}\gamma_m\,\ket{x_m}\bra{x_m},
\qquad \gamma_m\ge 0,
\label{eq:CAP_def}
\end{equation}
which modifies Eq.~\eqref{eq:eom_cm} by the replacement $e_0\to e_0-i\gamma_m$.
In the actual simulations, the chain length is $M=5000$ and the CAP is applied within an edge region of width $w=3000$ sites at both ends.
More explicitly, defining $|m|$ as the distance from the center, we set
\begin{equation}
\gamma_m=
\begin{cases}
0, & |m|\le M-w,\\
\gamma_0\left[\dfrac{|m|-(M-w)}{w}\right]^{p}, & |m|> M-w,
\end{cases}
\label{eq:CAP_profile}
\end{equation}
with $(\gamma_0,p)=(0.6,2)$.
This smooth polynomial ramp suppresses back-reflection of outgoing wavepackets and is a standard practical choice for minimizing spurious reflections in wavepacket propagation~\cite{manolopoulos2002derivation, riss1996investigation}.
We checked that, in the time windows used in the figures, the survival probabilities are insensitive to
moderate changes of $M$, $w$, and the CAP strength profile.
The coupled ordinary differential equations \eqref{eq:eom_a}--\eqref{eq:eom_cm} are integrated with a standard fourth-order Runge--Kutta scheme. We use a time step $\Delta t = 10^{-3}$, and integrate up to $t_{\max}=2.0\times 10^{4}$.
The initial conditions corresponding to $\ket{\psi(0)}=\ket{d_X}$ are $d_X(0)=1$, $d_{Y}(0)=0$, $c_m(0)=0$, where $X\ne Y$.
In each run with a fixed driving frequency $\omega$, the system is re-initialized before starting the time integration.

\bibliographystyle{jpsj}
\bibliography{refs}

\end{document}